\documentclass[runningheads,a4paper]{llncs}

\usepackage{amssymb}
\usepackage{graphicx,epsfig,epstopdf,endnotes,amsmath,multirow,hyperref,bm,threeparttable,subfigure,rotating,adjustbox}
\usepackage{url}
\usepackage{rotating}
\usepackage{setspace}
\usepackage{pifont,algorithm,algorithmicx,algpseudocode}

\urldef{\mailsa}\path|{luoyang, qingnishen, wuzh}@pku.edu.cn|
\newcommand{\keywords}[1]{\par\addvspace\baselineskip
\noindent\keywordname\enspace\ignorespaces#1}

\begin{document}
\begin{spacing}{0.93}

\mainmatter              
\title{PML: An Interpreter-Based Access Control Policy Language for Web Services}
%
\titlerunning{PML: An Interpreter-Based Access Control Policy Language..}  
%
\author{Yang Luo \and Qingni Shen \and Zhonghai Wu\thanks{Corresponding author}}
\authorrunning{Yang Luo et al.} 
%
\tocauthor{Yang Luo, Qingni Shen, and Zhonghai Wu}
\institute{Peking University, China,\\
\email{\{luoyang,qingnishen,wuzh\}@pku.edu.cn}}

%
%

\toctitle{PML}
\tocauthor{Authors' Instructions}
\maketitle

\begin{abstract}

Access control is an important component for web services such as a cloud. Current clouds tend to design the access control mechanism together with the policy language on their own. It leads to two issues: (i) a cloud user has to learn different policy languages to use multiple clouds, and (ii) a cloud service provider has to customize an authorization mechanism based on its business requirement, which brings high development cost. In this work, a new access control policy language called PERM modeling language (PML) is proposed to express various access control models such as access control list (ACL), role-based access control (RBAC) and attribute-based access control (ABAC), etc. PML's enforcement mechanism is designed in an interpreter-on-interpreter manner, which not only secures the authorization code with sandboxing, but also extends PML to all programming languages that support Lua. PML is already adopted by real-world projects such as Intel's RMD, VMware's Dispatch, Orange's Gobis and so on, which proves PML's usability. The performance evaluation on OpenStack, CloudStack and Amazon Web Services (AWS) shows PML's enforcement overhead per request is under 5.9$\mu$s.

\end{abstract}

\section{Introduction}

For the past few years, cloud computing has become a revolutionary power for enterprises to reduce their costs by using on-demand computational infrastructures \cite{takabi2010security}. As a public cloud provides its interface on Internet, performing access control on the cloud data serves as a critical part of the cloud's security. Over the last decades, a couple of access control policy languages have been proposed by the academic comunnity, like XACML \cite{oasis2016xacml}, SPL \cite{da2001spl}, Ponder \cite{damianou2001ponder}, etc. As one of the most representative security policy languages, OASIS's XACML is general-purpose authorization language based on the attribute-based access control (ABAC) \cite{yuan2005attributed} model. However, except scattered usage on JBoss, Axiomaics and OpenAZ, it is not widely accepted by the industrial world. Large public clouds like Amazon Web Services (AWS), Microsoft Azure and open-source clouds like OpenStack tend to design their own access control policy languages for their platforms. Similarly, network management platforms like HP Openview PolicyXpert and CiscoAssure also only support their own languages \cite{han2012survey}. For the past few years, there has been a considerable interest in environments that support multiple and complex access control policies with better expressiveness and functionality \cite{da2001spl,bertino1996supporting,carney1998comparison,jajodia1997unified,minsky1998unified}. However, a vast majority of them remain merely academic and lack practical acceptance owing to their complexity in usage or computation.

The diversity of access control between different clouds has imposed several challenges: 1) There are unique concepts and features for different access control models. e.g., role is defined in role-based access control (RBAC) \cite{sandhu1996role} and the attribute is defined in ABAC. Two security policy languages that use different access control models may have distinct semantics. For example, AWS Identity and Access Management (IAM) expresses RBAC and ABAC mixed authorization via abstractions like \texttt{Statement}, \texttt{Principle}, \texttt{Action} and \texttt{Resource}, while OpenStack uses totally different condition-action rule sets to express ABAC-like authorization. 2) The policy enforcement mechanism has to be implemented in a programming language. Different cloud platforms are probably developed in different languages, which causes that the policy checking logic for one cloud probably cannot be reused in another cloud. For example, OpenStack is developed in Python. CloudStack is developed in Java. AWS is developed in multiple languages: C++, Java, Python, etc. Therefore, even all these platforms support the same access control model, the corresponding policy enforcement mechanism still requires to be implemented on respective programming languages. The above challenges will cause two issues:


\begin{itemize}
\item[1.] If a cloud user deploys his business on multiple clouds, he has to learn different security policy languages. And when the cloud user migrates to another cloud, the original policy rules cannot be migrated and have to be rewritten in the policy language of the new cloud.
\item[2.] A cloud service provider (CSP) has to create a security policy language and the corresponding enforcement mechanism from scratch, which is a cost of development. Moreover, CSP's implementation may introduce security holes if it lacks adequate security expertise.
\end{itemize}

To solve the above issues, we design a security policy language called PERM modeling language (PML). PML is able to express a number of existing access control models like access control list (ACL) \cite{sandhu1994access}, RBAC, ABAC, etc. PML enforcement mechanism (PML-EM) is designed, which provides two features:

\begin{itemize}
  \item[1.] \textbf{Access Control Model Independent (Model-Ind)}: it enables cloud users to control the access to cloud data based on the same access control model across different clouds that support PML-EM. Moreover, a CSP can provide multiple access control models to its users without complicating the implementation of policy enforcement.
  \item[2.] \textbf{Implementation Language Independent (Impl-Ind)}: we design PML-EM in a cross-language manner. So a CSP does not even need to implement PML-EM itself, which lowers the cost of development. As the modern software advances much faster than before, even if the CSP is willing to implement PML-EM, it is hard to keep the same development pace for all PML-EM implementations of different programming languages, which can easily cause compatibility issues.
\end{itemize}

We implement PML-EM in Lua, a widely-used scripting language, which usually embeds its code in another programming language. Part of this work (without interpreter-on-interpreter) is implemented as an open-source project called Casbin \cite{casbin2018casbin}, which is hosted on GitHub. It has been already used in a number of projects such as Intel's RMD \cite{intel2018rmd}, VMware's Dispatch \cite{vmware2018dispatch}, Orange's Gobis \cite{orange2018gobis} in practice and recommended by Docker \cite{docker2018docker}. We compared the expressiveness of PML with existing policy languages and studied the case of Intel's RMD to show PML's usability. The performance evaluation on OpenStack, CloudStack and Amazon AWS show the policy enforcement delay introduced by PML-EM is less than 5.9$\mu$s on average with Lua JIT (just-in-time) enabled.


The contributions of this paper are as follows:


\begin{itemize}
\item[1.] PML, the new authorization policy language simplifies policy designing by separating the abstract authorization logic from the concrete policy rules. PML is not only expressive but also has good usability. PML is already put into practical use on a number of projects such as Intel's RMD, VMware's Dispatch, Orange's Gobis and so on. These facts prove PML's usability.
\item[2.] PML-EM, the enforcement mechanism of PML is implemented in Lua based on our interpreter-on-interpreter idea, which makes it directly usable for all programming languages that support Lua. A CSP can use it as a drop-in replacement for its original authorization mechanism without making efforts to implement it in a new language. The experimental results show PML-EM does not cause significant enforcement overhead.
\end{itemize}


The remainder of this paper is organized as follows. Section \ref{sec:related_work} describes related work. Section \ref{sec:pml} presents our PML language. Section \ref{sec:design} shows the design of our PML-EM enforcement framework. Section \ref{sec:evaluations} brings the experimental results. Section \ref{sec:conclusion} concludes the paper and points out our future work.

\section{Related Work}
\label{sec:related_work}

Over the past few decades, a wide variety of industrial and academic efforts have been done to provide a more secure cloud. Amazon AWS, currently the leading proprietary cloud, has supported identity-based authorization for cloud customers with its IAM service. Basic elements like users, groups and permissions are provided for customers to restrict the access to their own resources.


Access control as a service (ACaaS), proposed in \cite{wu2013acaas} by Wu et al., provided a new cloud service to provide comprehensive and fine-grained access control. It is claimed to support multiple access control models, whereas there is no evidence that this approach applies to the models except RBAC. And this work is highly based on IAM provided by AWS, which makes it difficult to apply for other clouds.

OpenStack access control (OSAC), proposed in \cite{tang2014extending} by Tang et al., has presented a formalized description for conceptions in Keystone, such as domains and tenants in addition to roles. It further proposed a domain trust extension for OSAC to facilitate secure cross-domain authorization. This work is orthogonal to ours, since it mainly focuses on the enhancement of Keystone. The domain trust decision made by OSAC can be used as an attribute in PML, which increases the granularity of the access control.

The work proposed in \cite{jin2014role} by Jin et al., has defined a formal ABAC specification suitable for infrastructure as a service (IaaS) and implemented it in OpenStack. It includes two models: the operational model $IaaS_{op}$ and the administrative model $IaaS_{ad}$, which provide fine-grained access control for tenants. However, this work does not support external functions or decision combination, which limits its flexibility to express a customized model.


Attribute-based access control (ABAC) \cite{hu2014sp} broadened the conception of roles into attributes compared to role-based access control (RBAC) \cite{sandhu1996role}. Extensible Access Control Markup Language (XACML), proposed in \cite{oasis2016xacml} by OASIS, is primarily based on the ABAC model by defining structural elements such as \texttt{Rule}, \texttt{Policy} and \texttt{Policy Set}. Combining algorithms like \texttt{deny-overrides} and \texttt{permit-overrides} are also provided to resolve conflicts between rules or policies. XACML has specified the whole architecture about the supporting entities like PEP, PDP, PIP and the exchanging structures between those entities.


Ponder, proposed in \cite{damianou2001ponder}, was a policy specification language for distributed systems. It supports access control by providing authorization, delegation, information filtering and refrain policies. Ponder also supports obligation policies that are event triggered condition-action rules for policy based management of networks and distributed systems. PML only supports authorization rules, but it uses expression evaluation to maximize the flexibility. Ponder can be extended by supporting external functions but it still lacks many features like the tenants, arithmetic and logical operators, etc. It supports static conflict detection but lacks decision combination (similar to conflict resolution).

Security policy language (SPL), proposed in \cite{da2001spl} by Ribeiro et al., supports the expression of entities, their relations and the comparison of properties and quantifiers from different policies. SPL lacks the external function support, which is a primary restriction for more flexible policy evaluation. And it also relies heavily on policy rules, which is unnecessary for some scenarios.


\section{PML}
\label{sec:pml}

When we talk about the access control policy, we do not quite distinguish between actual policy rules and enforcement logic which parses and executes these rules. A system that supports multiple security policies typically allows its users to design their own policy rules. However, the underlying enforcement logic of them is usually static and unchangeable, which is inflexible. To make the policy more customizable for cloud users, we extract the access control enforcement logic out of policy evaluation. As the enforcement logic usually reflects the underlying access control model that the policy rules belong to, it is a natural idea to separate the model from the policy rules. We call the abstract access control model which represents the enforcement logic as the ``model'' and the enforced policy rules as the ``policy'' or ``policy rules''. It is similar to the relationship of code and data in a program.

\begin{figure}[!t]
\centering
\includegraphics[width=2.5in]{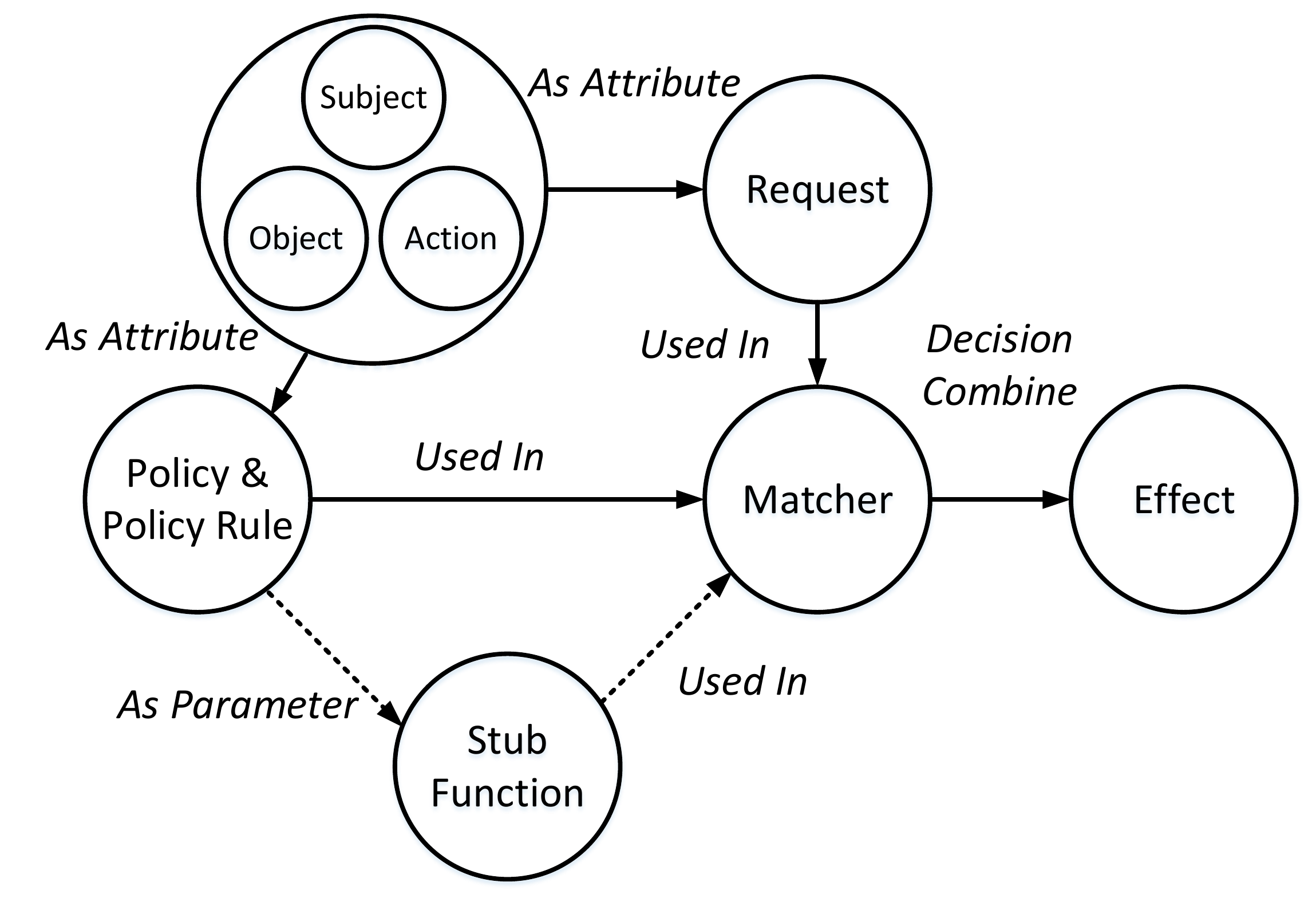}
\caption{The PERM metamodel.}
\label{fig_pml}
\end{figure}

In this section we propose the PERM modeling language (PML) which can be used to express an access control model. PML is based on the PERM metamodel, which is named after Policy-Effect-Request-Matcher. The PERM metamodel, as shown in Figure \ref{fig_pml}, has six basic primitives: \emph{Request}, \emph{Policy}, \emph{Policy Rule}, \emph{Matcher}, \emph{Effect} and \emph{Stub Function}.




\subsection{Primitives}

\textbf{\emph{Definition-1: Request}}, \emph{request} is defined as a key-value pair as follows:

\begin{equation}
\begin{split}
request &::= \textbf{r} : attributes \\
attributes &::= \{attr_1, attr_2, attr_3, ..\}
\end{split}
\end{equation}

The key is always \emph{r}, which represents the access request entity that needs to be mediated. The value is a list of attribute names that the request entity has. PML will parse the incoming request based on these attributes. An access request is usually represented by the classic triple: accessing entity (\texttt{sub}), accessed resource (\texttt{obj}) and the access method (\texttt{act}). In this condition, we have: $attributes = sub, obj, act$, or in PML's grammar: \texttt{r = sub, obj, act}

PML provides the flexibility for the user to customize his own request, like using $attributes = sub, act$ if the high-level authorization goal do not need to specify an particular resource, or $attributes =  tenant, sub, obj, act$ if the subject needs more information to be identified (e.g., the tenant a user belongs to).

\textbf{\emph{Definition-2: Policy}}, \emph{policy} is defined as a key-value pair as follows:

\begin{equation}
\begin{split}
policy &::= \textbf{p} : attributes \\
attributes &::= \{attr_1, attr_2, attr_3, ..\}
\end{split}
\end{equation}

The key is always \emph{p}, which represents an abstract policy rule entity. The value $attributes$ denotes the attribute names that \emph{p} has. A typical assignment is: $attributes = sub, obj, act$. It means the policy rule will have three fields: subject, object and action. In PML's grammar, it is: \texttt{p = sub, obj, act}


\textbf{\emph{Definition-3: Policy Rule}}, a policy rule is an instance of the above \emph{policy}. It is defined as tuple of values as follows:

\begin{equation}
policy\_rule ::= \{value_1, value_2, value_3, ..\}
\end{equation}

The number of elements in the tuple will be identical with the attribute names in \emph{policy}. When a policy rule is evaluated against a request, PML will assign the attribute names in \emph{policy} with the values in the policy rule in order.

For example, the above policy rule: \texttt{alice, data1, read} generates a binding to the attributes like: \texttt{p.sub = alice, p.obj = data1, p.act = read}.



\textbf{\emph{Definition-4: Matcher}}, a \emph{matcher} determines how the policy rules are evaluated against the request. It is defined as a boolean expression as follows:

\begin{equation}
\begin{split}
matcher &::= <boolean\_expr>(variables, constants, stub\_functions) \\
variables &::= \{\textbf{r}.attr_1, \textbf{r}.attr_2, .., \textbf{p}.attr_1, \textbf{p}.attr_2, ..\} \\
constants &::= \{const_1, const_2, const_3, ..\}
\end{split}
\end{equation}

$boolean\_expr$ connects $variables$ and $constants$ with operators. Supported operators in $<boolean\_expr>$ include arithmetic operators like $+$, $-$, $\times$, $\div$, relational operators like $==$ (equal), $!=$ (not equal), $>$, $<$ and logical operators like $\&\&$ (and), $||$ (or), $!$ (not) in the expression. The simplest matcher is: \texttt{m = r.sub == p.sub \&\& r.obj == p.obj \&\& r.act == p.act} (\texttt{m} represents the \texttt{matcher}). It means the matcher returns \texttt{true} only if subject, object and action in the access request exactly match the respective fields in a policy rule.

\textbf{\emph{Definition-5: Effect}}, \emph{effect} is defined as follows:

\begin{equation}
\begin{split}
e\hspace{-0.1em}f\hspace{-0.2em}fect &::= <boolean\_expr>(e\hspace{-0.1em}f\hspace{-0.2em}fect\_term_1, e\hspace{-0.1em}f\hspace{-0.2em}fect\_term_2, ..) \\
e\hspace{-0.1em}f\hspace{-0.2em}fect\_term &::= quantifier, condition \\
quantifier &::= some | any | max | min \\
condition &::= <expr>(variables, constants, stub\_functions) \\
variables &::= \{\textbf{r}.attr_1, \textbf{r}.attr_2, .., \textbf{p}.attr_1, \textbf{p}.attr_2, ..\} \\
constants &::= \{const_1, const_2, const_3, ..\}
\end{split}
\end{equation}

The \emph{effect} primitive determines whether the request should be approved if multiple policy rules match the request. In other words, it can form a single decision by combining multiple decision results from the matched policy rules. It is comprised of multiple $effect\_term$ in the boolean expression. In a $effect\_term$, the $quantifier$ aggregates the multiple decisions from the valid set for $condition$ into a single boolean value. It can be $some$, $max$ or $min$. $condition$ functions in a similar way as $matcher$, but it is used for filtering valid decisions instead of matching the policy rules with the request.

\begin{itemize}
\item $quantifier = some$: if there exists one policy rule that satisfied both the matcher and $condition$ (aka made the matcher and $condition$ evaluated as \texttt{true}), use the $effect$ attribute value of the policy rule as the final value. If the $effect$ attribute is not specified, it is \texttt{true} by default.
\item $quantifier = max$: if there are multiple policy rules that satisfied the matcher, the final value will be the $effect$ attribute value of the policy rule that makes $condition$ maximized.
\item $quantifier = min$: if there are multiple policy rules that satisfied the matcher, the final value will be the $effect$ attribute value of the policy rule that makes $condition$ minimized.
\end{itemize}


The simplest allow-override $effect$ is: \texttt{e = some(where (p.eft == allow))} (\texttt{e} represents the \texttt{effect}, \texttt{eft} is the attribute name for a matched policy rule \texttt{p}, \texttt{where} is a keyword). The whole policy effect means if there is any matched policy rule of \texttt{allow}, the final effect is \texttt{allow} (aka allow-override). Similarly, deny-override can be expressed as: \texttt{e = !some(where (p.eft == deny))}. It means the request is allowed only when there is no matched policy rules with \texttt{deny} as the value for the effect attribute (aka deny-override). The $effect$ can even be connected with logical expressions: \texttt{e = some(where (p.eft == allow)) \&\& !some(where (p.eft == deny))}. It means the request is authorized only when at least one matched policy rule of \texttt{allow}, and there is no matched policy rule of \texttt{deny}. Therefore, in this way both the allow and deny authorizations are supported, and \texttt{deny} overrides.

\textbf{\emph{Definition-6: Stub Function}}, \emph{stub function} refers to functions that can be customized by the policy designer. It is defined as follows:

\begin{equation}
\begin{split}
stub\_function &::= function\_name : parameters \\
parameters &::= \{param_1, param_2, param_3, ..\}
\end{split}
\end{equation}

Although PML is expressive by supporting arithmetic and logical operators, there is still a lot of complicated evaluation logic which is difficult to describe, or the permission checking needs to query an external source like a database. The \emph{stub function} enables the policy designer to write his own customized logic. A \texttt{stub function} can be used in a \texttt{matcher} or \texttt{effect}. It can return any type of values such as boolean, integer or string, which will be evaluated into the final decision. The stub function can be realized in the same language as the system that uses the authorization. As an example, if we have a function called \texttt{obj\_match} to match the object in the request with the object specified in policy rules, we can register it as: $register\_function("obj\_match", obj\_match)$, then we can specify it in the matcher.


\subsection{Extended Concepts}

\textbf{\emph{Definition-7: Has\_Role}}, in PML, we do not directly define the \emph{role} concept. Instead, we define the stub function that determines whether there is a ``role inheritance relationship'' between two attributes. It is defined as follows:

\begin{equation}
\begin{split}
has\_role &::= function\_name : parameters \\
function\_name &::= ``g" \\
parameters &::= \{attr_1, attr_2\}
\end{split}
\end{equation}

A common usage of the \texttt{has\_role} function is shown as follows.

{\tt
\small
\begin{quote}
\begin{verbatim}
m = g(r.sub, p.sub) && r.obj == p.obj && r.act == p.act
\end{verbatim}
\end{quote}
}

The matcher checks whether the subject in the policy rule is the role of the subject in the request. When realizing \texttt{has\_role}, the policy designer can store the role hierarchy externally and check the role inheritance relationship between the two attributes in the function. Therefore, By using \texttt{has\_role} stub function, PML proves to support the classic RBAC1 (RBAC with role hierarchy) scenario.

\textbf{\emph{Definition-8: Has\_Tenant\_Role}}, to express the tenant in the cloud scenario, PML defined a stub function called $has\_tenant\_role$ to determine whether there is a ``role inheritance relationship'' between two attributes inside a tenant as follows:

\begin{equation}
\begin{split}
has\_tenant\_role &::= function\_name : parameters \\
function\_name &::= ``g" \\
parameters &::= \{attr_1, attr_2, tenant\}
\end{split}
\end{equation}

In other words, the RBAC roles in PML can be global or tenant-specific. Tenant-specific roles mean that the roles for a user can be different when the user is in different tenants. The following policy rules and role assignments show such a scenario: the users with \texttt{admin} role can use and manage any resources in the tenant, but the users with \texttt{user} role can only use the resources inside the tenant. Alice has \texttt{admin} role in \texttt{tenant1}, and has \texttt{user} role in \texttt{tenant2}. So she can use and manage resources in \texttt{tenant1}. However, as Alice is not an \texttt{admin} in \texttt{tenant2}, she cannot manage resources in that tenant.

{\tt
\small
\begin{quote}
\begin{verbatim}
Alice has the admin role in tenant1
Alice has the user role in tenant2
admin, *, (use|manage)
user, *, use
\end{verbatim}
\end{quote}
}

We can design the matcher logic as follows. It first uses $has\_tenant\_role$ stub function to determine whether the requesting subject has the role specified in the policy rules in the tenant specified by the requested object's tenant. Then the matcher compares the request's object with the policy rule's object. We used additional: \texttt{|| p.obj == "*"} logic to allow to specify \texttt{*} wildcard (which means ``all'') in the policy rule. We also design a $regex\_match$ function to allow to use regular expressions inside the policy rules. \texttt{(use|manage)} means to either match \texttt{use} or \texttt{manage} in the action.

{\tt
\small
\begin{quote}
\begin{verbatim}
m = has_tenant_role(r.sub, p.sub, r.obj.tenant) && (r.obj == p.obj ||
p.obj == "*") && regex_match(r.act, p.act)
\end{verbatim}
\end{quote}
}

%
%


\section{Policy Enforcement}
\label{sec:design}

In this section, we designed the PML interpreter as the enforcement mechanism of the PML policy. It supports 3 primitives: \texttt{load model}, \texttt{load policy} and \texttt{evaluate}. The primitives \texttt{load model} and \texttt{load policy} simply load the PML model and rules from the database in the initial stage. In \texttt{evaluate}, the variables of request, policy rules and stub functions are substituted into the matcher. Then the Lua interpreter evaluates the matcher to get the boolean decision for the access request. Figure \ref{fig_sandbox} shows the work flow of PML-EM, including how PML interpreter evaluates a request, and how PML model and policy rules are designed and loaded.


%
%

\subsection{Interpreter-on-Interpreter}

A security issue here is that the stub functions in PML allow to run user-side code during the policy enforcement. It may cause the arbitrary code execution attack. Possible measures to solve this issue include:

\begin{itemize}
\item[1.] Run stub functions in a sandbox. The code in a stub function cannot do harm to the outer PML interpreter or the whole system.
\item[2.] Perform static code inspection when a stub function is registered in CSP. The malicious code can be detected during the inspection.
\item[3.] Provide a user interface for function editing, which supplies predefined building blocks to establish a stub function.
\end{itemize}

\begin{figure}[!t]
\centering
\includegraphics[width=3.0in]{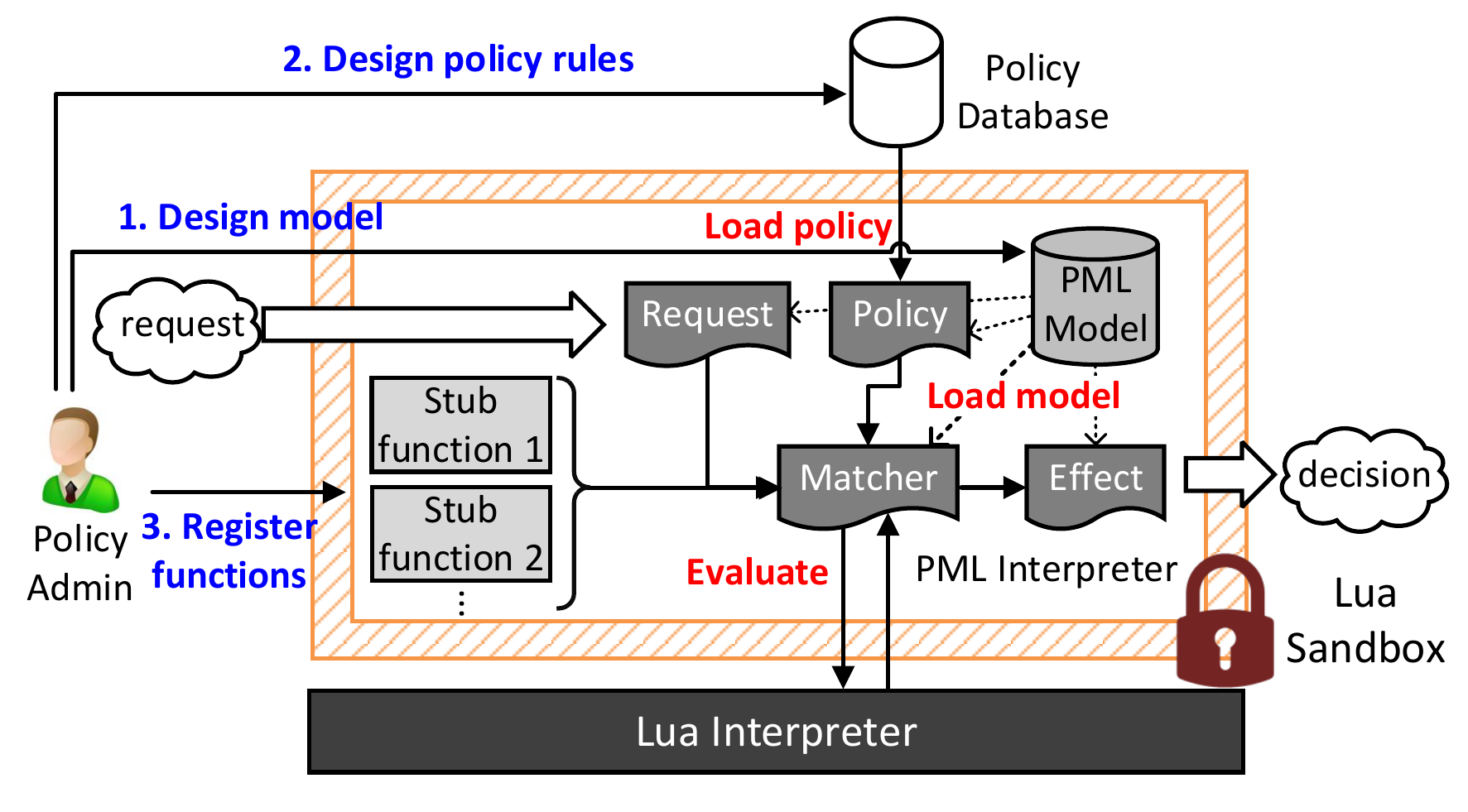}
\caption{The work flow of PML-EM.}
\label{fig_sandbox}
\end{figure}

Unfortunately, all above ways require additional mechanisms, which is inconvenient. Here we come up with a painless solution: as shown in Figure \ref{fig_sandbox}, we write the PML interpreter in Lua, which means the PML interpreter runs itself on a Lua interpreter. We call it an \textbf{interpreter-on-interpreter} (IoI) design. As far as we know, we are the first to use the IoI approach in the security policy enforcement field. Fortunately, most popular programming languages have a mature implementation of the Lua interpreter. And it is rather small in size (mostly less than 1000 KB) and can be easily embedded in the underlying systems that needs to be access controlled. In this way, the cloud platform can use PML-EM as a drop-in replacement for the authorization and does not need to implement any parts of it. This answers the question about how Impl-Ind is achieved. On the whole, the advantages of IoI are two fold:

\begin{itemize}
\item[1.] \textbf{Cross-platform}: IoI enabled PML-EM to be easily deployed on systems written in different languages without causing extra development efforts for CSPs. Moreover, maintaining a single implementation also minimizes the attack surfaces and inconsistency between different branches.
\item[2.] \textbf{Sandboxing}: as stated above, the execution of stub functions should be restricted in a secure environment. We find Lua can provide a natural sandbox for it: we only import a subset of Lua standard libraries: \texttt{base}, \texttt{table}, \texttt{string}, and \texttt{math} in the Lua interpreter. Operating system related libraries such as \texttt{os} are not imported. The stub functions are restricted to be written in Lua. And the Lua code cannot escape like invoking system calls without the library \texttt{os}. In this way, we managed to restrict what a stub function can do.
\end{itemize}

\subsection{Interpreting Overhead}


Some may doubt that IoI will result in poor performance on policy enforcement, as it introduces additional interpreter layer in the PML-EM stack. It may be true for most scenarios but not for PML-EM because of two reasons:

1. Because of the design of IoI, there are two interpreters in PML-EM: the PML interpreter and the Lua interpreter. The PML matcher is evaluated as an expression in the PML interpreter. And the PML interpreter is written in Lua and runs in a Lua interpreter. To avoid two layers of interpretation, IoI uses an optimization technique: the PML matcher follows nearly the same grammar as Lua. The PML interpreter directly uses the Lua interpreter to execute the PML matcher instead of interpreting and evaluating the matcher by itself, as shown in Figure \ref{fig_sandbox}. It is similar to the hardware-assisted virtualization technique \cite{fisher2006hardware} in the virtualization area. Hardware-assisted virtualization runs the virtual machine code directly in a physical processor core instead of using a binary emulator. It avoids the binary translation overhead and greatly increases the performance. So theoretically, the overhead of both PML interpreter and Lua interpreter is very close to that of a single Lua interpreter.

2. Network-based systems, especially public clouds can tolerate delays resulting from IoI, compared to the round trip time (RTT) of a request on the Internet. A local system, such as a typical operating system, may not be suitable to use a interpreter-based mechanism like IoI.

Based on the above two arguments, the interpreter layer of PML-EM will not cause any significant overhead. We will evaluate the performance of PML-EM more detailedly in Section \ref{sec:evaluations_performance}.



\section{Evaluations}
\label{sec:evaluations}

\subsection{Usability}

Part of this work (without interpreter-on-interpreter) is implemented as an open-source project called Casbin, which is hosted on GitHub. It has been already used in a number of projects such as Intel's RMD, VMware's Dispatch, Orange's Gobis in practice and recommended by Docker. We will use Intel RMD as an example and show how PML simplifies the authorization for these projects.

\subsubsection{Intel RMD}

Intel Resource Management Daemon (RMD) provides a central interface for hardware resource management tasks on Intel's x86 platforms. RMD offers RESTFul API for its users to manage the shared resources such as CPU cache and memory bandwidth. It uses PML for the access control to its RESTful calls. The PML model is:

{\tt
\small
\begin{quote}
\begin{verbatim}
r = sub, obj, act
p = sub, obj, act
g = _, _
m = g(r.sub, p.sub) && key_match(r.obj, p.obj) && regex_match(r.act,
p.act)
e = some(where (p.eft == allow))
\end{verbatim}
\end{quote}
}

The PML policy rules are as follows. RMD has two roles: \texttt{user} and \texttt{root}. \texttt{root} inherits all the permissions the \texttt{user} role has. \texttt{user} can read all resources, such as workloads, cache, policy, etc. But only \texttt{root} can modify or delete the workloads. The \texttt{root} role is assigned to the \texttt{admin} user. Overall, this is not a very complicated case, but it shows the usability of PML on real-world projects.


{\tt
\scriptsize
\begin{quote}
\begin{verbatim}
p, user, /cache/l*/*, GET
p, user, /cache/l*, GET
p, user, /cache, GET
p, user, /policy, GET
p, user, /workloads, GET
p, user, /workloads/*, GET
p, user, /hospitality, GET

p, root, /workloads, POST
p, root, /workloads/*, (PATCH)|(DELETE)

g, root, user
g, admin, root
\end{verbatim}
\end{quote}
}

\subsubsection{Expressiveness}

\begin{table*}
\caption{Expressiveness of different security policy languages}
\scriptsize
\center
\begin{threeparttable}
\renewcommand{\arraystretch}{1.3}
  \label{table_expressiveness}
  \begin{tabular}{ccccccccccccc} \hline
    {Policy} & \multicolumn{11}{c}{Supported model \& feature}\\
    \cline{2-12}
    language & ACL & BLP & RBAC & ABAC & SoMP\tnote{1} & MT\tnote{2} & RESTful & DC\tnote{3} & SF\tnote{4} & RH\tnote{5} & Impl-Ind\\ \hline
    OpenStack & \ding{55} & \ding{55} & \ding{51} & \ding{51} & \ding{55} & \ding{51} & \ding{51} & \ding{55} & \ding{55} & \ding{55} & \ding{55}\\ 
    AWS IAM & \ding{51} & \ding{55} & \ding{51} & \ding{51} & \ding{55} & \ding{51} & \ding{55} & \ding{51} & \ding{55} & \ding{55} & \ding{55}\\ 
    ACaaS \cite{wu2013acaas} & \ding{55} & \ding{55} & \ding{51} & \ding{55} & \ding{55} & \ding{51} & \ding{51} & \ding{51} & \ding{55} & \ding{55} & \ding{55}\\ 
    OSAC \cite{tang2014extending} & \ding{55} & \ding{55} & \ding{51} & \ding{55} & \ding{55} & \ding{51} & \ding{51} & \ding{55} & \ding{55} & \ding{55} & \ding{55}\\ 
    $IaaS_{op}$ \cite{jin2014role} & \ding{51} & \ding{55} & \ding{51} & \ding{51} & \ding{55} & \ding{51} & \ding{51} & \ding{55} & \ding{55} & \ding{55} & \ding{55}\\ 
    XACML \cite{oasis2016xacml} & \ding{51} & \ding{51} & \ding{51} & \ding{51} & \ding{55} & \ding{51} & \ding{51} & \ding{51} & \ding{51} & \ding{55} & \ding{55}\\ 
    Ponder \cite{damianou2001ponder} & \ding{55} & \ding{55} & \ding{51} & \ding{51} & \ding{55} & \ding{55} & \ding{55} & \ding{55} & \ding{51} & \ding{51} & \ding{55}\\ 
    SPL \cite{da2001spl} & \ding{51} & \ding{55} & \ding{51} & \ding{51} & \ding{55} & \ding{55} & \ding{55} & \ding{51} & \ding{55} & \ding{55} & \ding{55}\\ 
    PML (ours) & \ding{51} & \ding{51} & \ding{51} & \ding{51} & \ding{51} & \ding{51} & \ding{51} & \ding{51} & \ding{51} & \ding{51} & \ding{51}\\ \hline
\end{tabular}
\begin{tablenotes}
  \scriptsize
  \item[1] Separation of model and policy (SoMP): separates the description for the access control model from concrete policy rules.
  \item[2] Multi-tenancy (MT): allows a cloud-based system to support isolated permissions for multiple tenants.
  \item[3] Decision combination (DC): can form a single decision from all the policies that apply to an access request.
  \item[4] Stub function (SF): if external logic is supported.
  \item[5] Role hierarchy (RH): supports the RBAC role hierarchy or not.
\end{tablenotes}
\end{threeparttable}
\end{table*}

Moreover, as shown in Table \ref{table_expressiveness}, we compared PML with other popular policy languages such as XACML, Ponder, SPL and so on in expressiveness. The object owner in an ACL policy can be expressed by an \texttt{owner} attribute in PML. Similarly, subject's and object's security levels in the Bell-LaPadula (BLP) model \cite{bell1973secure} can be expressed by a \texttt{level} attribute in PML. RBAC is supported by all the languages. And most of them support ABAC. Multi-tenancy is a fundamental feature for a cloud, which is supported by several languages. XACML is a relatively expressive language, as it supports most of the features. But it does not separate the authorization logic from concrete policy rules. It does not mean this separation is definitely a better design, but we believe for a scenario that needs complex authorization logic and large number of rules, PML's modular design helps to simplify the problem. More details for these languages have been elaborated in Section \ref{sec:related_work}.

%
%
%
%
%
%

\subsection{Performance}
\label{sec:evaluations_performance}

In order to evaluate the performance under the same conditions, we manually designed PML policies that are equivalent to the original policies of OpenStack, CloudStack and AWS IAM in advance. It guarantees the compared authorization mechanisms to generate the same decision for an access request. Due to limited space, we only show how we setup PML policy for OpenStack. OpenStack provides a simple security policy language based on JSON. We first show a portion of Nova built-in policy rules:


{\tt
\small
\begin{quote}
\begin{verbatim}
{"admin_or_owner": "is_admin:True or project_id:%(project_id)s",
 "default": "rule:admin_or_owner",
 "compute:get": "",
 "compute:get_all_tenants": "is_admin:True",
 "compute:start": "rule:admin_or_owner"}
\end{verbatim}
\end{quote}
}

An OpenStack policy rule uses an action-condition format, which means only when the \texttt{condition} is met, then the access of \texttt{action} will be permitted. \texttt{compute:get} is an action to retrieve a virtual machine instance in OpenStack. \texttt{condition} can use property comparison. For example, \texttt{is\_admin:True} means comparing \texttt{is\_admin} property of the subject with \texttt{True}. \texttt{project\_id:\%(project\_
id)s} means comparing \texttt{project\_id} of the subject with \texttt{project\_id} of the object. \texttt{and}, \texttt{or} can be used as connectors. \texttt{rule:admin\_or\_owner} is a special condition. It means to include an already defined rule named \texttt{admin\_or\_owner}. Based on the above analysis, the resulting PML model is like:

{\tt
\small
\begin{quote}
\begin{verbatim}
r = sub, obj, act
p = act
m = r.sub.role == "admin" || r.sub.is_admin == true || (r.act == p.act
&& r.sub.project_id == r.obj.project_id)
e = some(where (p.eft == allow))
\end{verbatim}
\end{quote}
}

The policy rules are:

{\tt
\small
\begin{quote}
\begin{verbatim}
compute:get
compute:get_all_tenants
compute:start
\end{verbatim}
\end{quote}
}

\begin{table*}
\scriptsize
\center
  \caption{Benchmarks of policy enforcement overhead for: i) original mechanism, ii) non-IoI, iii) Lua, iv) Lua with JIT}
  \label{table_performance}
  \begin{tabular}{cccccccc} \hline
    \multirow{2}{*}{Benchmark} & \multirow{2}{*}{Lang} & Original & Non-IoI & \multicolumn{2}{c}{Lua} & \multicolumn{2}{c}{Lua with JIT}\\
    \cline{5-8}
     & & overhead/ms & overhead/ms & Interpreter & Overhead/ms & Interpreter & Overhead/ms\\ \hline
    Tempest & Python & 7110.09 & 15186.35 (2x) & Lupa & 50472.33 (7x) & Lupa & 2261.59 (0.32x)\\ 
    CloudStack & Java & 15.60 & N/A & LuaJava & 398.61 (25x) & Nonlua & 17.77 (1.14x) \\ 
    AWS IAM & C++ & 1.53 & N/A & LuaJIT & 25.41 (16x) & LuaJIT & 1.64 (1.07x) \\ \hline
\end{tabular}
\end{table*}

\begin{table}
\scriptsize
\center
  \caption{Benchmarks of delay per request for original mechanism \& PML-EM (JIT)}
  \label{table_performance_pt}
  \begin{tabular}{ccccc} \hline
    \multirow{2}{*}{Benchmark} & \multirow{2}{*}{\#Tests} & \multicolumn{3}{c}{Delay per request ($\mu$s)}\\
    \cline{3-5}
     &  & Original & PML-EM & \%Incr\\ \hline
    Nova & 737 & 472.80 & 174.90 & -63.0\\ 
    Glance & 270 & 403.32 & 139.74 & -65.4\\ 
    Neutron & 738 & 395.73 & 164.01 & -58.6\\ 
    Cinder & 149 & 457.20 & 133.27 & -70.9\\ 
    Heat & 397 & 371.63 & 73.69 & -80.2\\ 
    Ceilometer & 656 & 311.97 & 81.81 & -73.8\\ 
    CloudStack & 374 & 41.71 & 47.52 & 13.9\\ 
    AWS IAM & 261 & 5.84 & 6.27 & 7.3\\ \hline
\end{tabular}
\end{table}

To improve the efficiency, we used the JIT technique to compile Lua into machine code. For simplicity, all the policy rules are stored in the memory. So the performance overhead does not involve any policy loading time. The benchmarking result is shown in Table \ref{table_performance}. There are totally 4 groups of results: the \texttt{Original Mechanism}, \texttt{Non-IoI}, \texttt{Lua} and \texttt{Lua with JIT}. \texttt{Original Mechanism} means the vanilla policy enforcement mechanism for that language. \texttt{Non-IoI} is the PML-EM without Lua interpreter. Because of the additional PML interpreter, the enforcement overhead has enlarged by about 2x. It is notable that in non-IoI, we implement PML interpreter in Python's \texttt{eval} function instead of using Lua. \texttt{eval} dynamically interprets and executes the plain text (PML matcher in our case) as Python code, which dramatically affects the performance. Owing to the development cost, we only implement PML-EM in Python to compare with the original OpenStack's mechanism. We believe the results will be similar for Java and C++.

The \texttt{Lua} column in Table \ref{table_performance} represents the IoI-enabled PML-EM. It shows that PML-EM can be directly used to enforce different policies such as OpenStack, CloudStack or IAM for systems written in different languages (Python, Java or C++). It also proves PML-EM's advantages we claimed: Model-Ind and Impl-Ind. We can observe the policy enforcement overhead has greatly increased owing to the additional layer of Lua interpreter: 7x for Tempest, 25x for CloudStack and 16x for AWS IAM. It is understandable: as a scripting language, Lua code runs slower than bytecode-based Java and compiled language C++. Python is a scripting language similar to Lua. So introducing a Lua interpreter will influence the performance for Java and C++ systems more than Python. The \texttt{Lua with JIT} column in Table \ref{table_performance} shows the overhead of PML-EM with JIT-enabled Lua interpreter. JIT compiles Lua code (including PML interpreter and PML matcher) into machine code, so it dramatically improves the efficiency compared with non-JIT. It is notable that Lupa in the Tempest benchmark compiles Lua into machine code, which is faster than Python. It causes PML-EM to have a smaller overhead (0.32x) than the original policy enforcement mechanism.

Table \ref{table_performance_pt} shows the policy enforcement overhead per request. We can see that the final delay of PML-EM with JIT for OpenStack has decreased by at least 50\%. This is because the compiled Lua code is much faster than Python. The maximum delay happens in the CloudStack benchmark, because CloudStack policy is more complicated than the others and requires more enforcement efforts. We can also find the PML-EM enforcement delays for all benchmarks are less than 5.9$\mu$s. Even if we do not consider the request processing time in the cloud, it still causes no more than 0.2\% difference compared to the normal transmission delay of the Internet (30-100ms RTT from ping response time).

A seeming drawback of using JIT is that all Lua code requires to be compiled before running. The Lua code in PML-EM includes the PML interpreter and PML matcher. The PML interpreter contains the main logic of PML-EM and is not a frequently changed part, unless when PML-EM is upgraded. The PML matcher is part of PML policy, so it needs to be re-compiled when the policy administrator modifies the policy. However, policy renewal is usually a manual operation and far less frequent than the policy enforcement performed on each access. So it is acceptable on performance to use JIT to compile PML matcher into machine code.

\section{Conclusion and Future Work}
\label{sec:conclusion}

As far as we know, PML is the first access control policy language that separates the authorization logic from concrete policy rules. This design enables PML to express models like ACL, RBAC, ABAC without being too heavy-weight. As PML's enforcement mechanism, PML-EM is implemented in Lua based on the new interpreter-on-interpreter idea. It has two features: Model-Ind and Impl-Ind. Model-Ind means different access control models can be supported by PML-EM. Impl-Ind means PML-EM can run on systems written in all programming languages that support Lua. In PML-EM, We have designed the IoI mechanism which provides cross-platform and sandboxing. We show PML's usability in real-world projects like Intel's RMD. We also analyze the PML's evaluation performance on cloud platforms such as OpenStack, CloudStack and AWS IAM. When Lua JIT is enabled, the enforcement overhead per request is under 5.9$\mu$. Our future work is to establish a public authorization policy sharing platform like GitHub for researchers and security engineers to share their access control ideas and examples with others.




\section*{Acknowledgment}

The authors sincerely thank anonymous reviewers for their valuable suggestions and comments. This research was supported by the National Natural Science Foundation of China under Grant No. 61672062.

{
\bibliographystyle{IEEEtran}
\bibliography{PML}}

\begin{thebibliography}{10}
\providecommand{\url}[1]{#1}
\csname url@samestyle\endcsname
\providecommand{\newblock}{\relax}
\providecommand{\bibinfo}[2]{#2}
\providecommand{\BIBentrySTDinterwordspacing}{\spaceskip=0pt\relax}
\providecommand{\BIBentryALTinterwordstretchfactor}{4}
\providecommand{\BIBentryALTinterwordspacing}{\spaceskip=\fontdimen2\font plus
\BIBentryALTinterwordstretchfactor\fontdimen3\font minus
  \fontdimen4\font\relax}
\providecommand{\BIBforeignlanguage}[2]{{%
\expandafter\ifx\csname l@#1\endcsname\relax
\typeout{** WARNING: IEEEtran.bst: No hyphenation pattern has been}%
\typeout{** loaded for the language `#1'. Using the pattern for}%
\typeout{** the default language instead.}%
\else
\language=\csname l@#1\endcsname
\fi
#2}}
\providecommand{\BIBdecl}{\relax}
\BIBdecl

\bibitem{takabi2010security}
H.~Takabi, J.~B. Joshi, and G.-J. Ahn, ``Security and privacy challenges in
  cloud computing environments,'' \emph{IEEE Security \& Privacy}, no.~6, pp.
  24--31, 2010.

\bibitem{oasis2016xacml}
OASIS, ``extensible access control markup language (xacml) version 3.0,''
  \url{http://docs.oasis-open.org/xacml/3.0/xacml-3.0-core-spec-os-en.html},
  2016.

\bibitem{da2001spl}
C.~da~Cruz~Ribeiro, A.~d. C.~M. Z{\'u}quete, P.~Ferreira, and P.~Guedes, ``Spl:
  An access control language for security policies with complex constraints,''
  in \emph{Network and Distributed System Security Symposium (NDSS¡¯01)}, pp.
  89--107.

\bibitem{damianou2001ponder}
N.~Damianou, N.~Dulay, E.~Lupu, and M.~Sloman, ``The ponder policy
  specification language,'' in \emph{Policies for Distributed Systems and
  Networks}.\hskip 1em plus 0.5em minus 0.4em\relax Springer, 2001, pp. 18--38.

\bibitem{yuan2005attributed}
E.~Yuan and J.~Tong, ``Attributed based access control (abac) for web
  services,'' in \emph{Web Services, 2005. ICWS 2005. Proceedings. 2005 IEEE
  International Conference on}.\hskip 1em plus 0.5em minus 0.4em\relax IEEE,
  2005.

\bibitem{han2012survey}
W.~Han and C.~Lei, ``A survey on policy languages in network and security
  management,'' \emph{Computer Networks}, vol.~56, no.~1, pp. 477--489, 2012.

\bibitem{bertino1996supporting}
E.~Bertino, S.~Jajodia, and P.~Samarati, ``Supporting multiple access control
  policies in database systems,'' in \emph{Security and Privacy, 1996.
  Proceedings., 1996 IEEE Symposium on}.\hskip 1em plus 0.5em minus 0.4em\relax
  IEEE, 1996, pp. 94--107.

\bibitem{carney1998comparison}
M.~Carney and B.~Loe, ``A comparison of methods for implementing adaptive
  security policies,'' in \emph{Proceedings of the Seventh USENIX Security
  Symposium}, 1998, pp. 1--14.

\bibitem{jajodia1997unified}
S.~Jajodia, P.~Samarati, V.~Subrahmanian, and E.~Bertino, ``A unified framework
  for enforcing multiple access control policies,'' in \emph{ACM Sigmod
  Record}, vol.~26, no.~2.\hskip 1em plus 0.5em minus 0.4em\relax ACM, 1997,
  pp. 474--485.

\bibitem{minsky1998unified}
N.~H. Minsky and V.~Ungureanu, ``Unified support for heterogeneous security
  policies in distributed systems,'' in \emph{7th USENIX Security Symposium},
  1998, pp. 131--142.

\bibitem{sandhu1996role}
R.~S. Sandhu, E.~J. Coyne, H.~L. Feinstein, and C.~E. Youman, ``Role-based
  access control models,'' \emph{Computer}, no.~2, pp. 38--47, 1996.

\bibitem{sandhu1994access}
R.~S. Sandhu and P.~Samarati, ``Access control: principle and practice,''
  \emph{IEEE communications magazine}, vol.~32, no.~9, pp. 40--48, 1994.

\bibitem{casbin2018casbin}
Y.~Luo, ``Casbin: an authorization library that supports access control models
  like acl, rbac, abac in golang,'' \url{https://github.com/casbin/casbin},
  2018.

\bibitem{intel2018rmd}
Intel, ``Intel rmd: a central uniform interface portal for hardware resource
  management tasks on x86 platforms,'' \url{https://github.com/intel/rmd},
  2018.

\bibitem{vmware2018dispatch}
VMware, ``Vmware dispatch: a framework for deploying and managing serverless
  style applications,'' \url{https://github.com/vmware/dispatch}, 2018.

\bibitem{orange2018gobis}
Orange, ``Orange gobis: a lightweight api gateway which can be used as a
  standalone server.'' \url{https://github.com/orange-cloudfoundry/gobis},
  2018.

\bibitem{docker2018docker}
Docker, ``Docker documentation: Use docker engine plugins.''
  \url{https://docs.docker.com/engine/extend/legacy_plugins/#authorization-plugins},
  2018.

\bibitem{wu2013acaas}
R.~Wu, X.~Zhang, G.-J. Ahn, H.~Sharifi, and H.~Xie, ``Acaas: Access control as
  a service for iaas cloud,'' in \emph{Social Computing (SocialCom), 2013
  International Conference on}.\hskip 1em plus 0.5em minus 0.4em\relax IEEE,
  2013, pp. 423--428.

\bibitem{tang2014extending}
B.~Tang and R.~Sandhu, ``Extending openstack access control with domain
  trust,'' in \emph{Network and System Security}.\hskip 1em plus 0.5em minus
  0.4em\relax Springer, 2014, pp. 54--69.

\bibitem{jin2014role}
X.~Jin, R.~Krishnan, and R.~Sandhu, ``Role and attribute based collaborative
  administration of intra-tenant cloud iaas,'' in \emph{Collaborative
  Computing: Networking, Applications and Worksharing (CollaborateCom), 2014
  International Conference on}.\hskip 1em plus 0.5em minus 0.4em\relax IEEE,
  2014, pp. 261--274.

\bibitem{hu2014sp}
V.~Hu, D.~Ferraiolo, R.~Kuhn, A.~Schnitzer, K.~Sandlin, and K.~Scarfone, ``Sp
  800-162. guide to attribute based access control (abac) definitions and
  considerations,'' \emph{Tech. Rep.}, 2014.

\bibitem{fisher2006hardware}
J.~Fisher-Ogden, ``Hardware support for efficient virtualization,''
  \emph{University of California, San Diego, Tech. Rep}, 2006.

\bibitem{bell1973secure}
D.~E. Bell and L.~J. LaPadula, ``Secure computer systems: Mathematical
  foundations,'' DTIC Document, Tech. Rep., 1973.

\end{thebibliography}

\end{spacing}
\end{document}